\documentclass[conference]{IEEEtran}
\IEEEoverridecommandlockouts
\usepackage{cite}
\usepackage{amsmath,amssymb,amsfonts}
\usepackage{algorithmic}
\usepackage{graphicx}
\usepackage{textcomp}
\usepackage{xcolor}
\usepackage{color}
\usepackage{multirow}
\graphicspath{{figures/}}
\def\BibTeX{{\rm B\kern-.05em{\sc i\kern-.025em b}\kern-.08em
    T\kern-.1667em\lower.7ex\hbox{E}\kern-.125emX}}
\begin{document}

\title{D2-LRR: A Dual-Decomposed MDLatLRR Approach for Medical Image Fusion\\
\thanks{This work was supported by the National Natural Science Foundation of China (62202205, 62020106012), the Fundamental Research Funds for the Central Universities (JUSRP123030).}
}

\author{
\IEEEauthorblockN{1\textsuperscript{st} Xu Song}
\IEEEauthorblockA{\textit{School of Artificial Intelligence and Computer Science} \\
\textit{Jiangnan University}\\
Wuxi, China \\
xu\_song\_jnu@163.com}
\and
\IEEEauthorblockN{2\textsuperscript{nd} Tianyu Shen}
\IEEEauthorblockA{\textit{School of Artificial Intelligence and Computer Science} \\
\textit{Jiangnan University}\\
Wuxi, China \\
tianyu\_shen\_jnu@163.com}
\and
\IEEEauthorblockN{3\textsuperscript{rd} Hui Li*}
\IEEEauthorblockA{\textit{School of Artificial Intelligence and Computer Science} \\
\textit{Jiangnan University}\\
Wuxi, China \\
lihui.cv@jiangnan.edu.cn}
\and
\IEEEauthorblockN{4\textsuperscript{th} Xiao-Jun Wu}
\IEEEauthorblockA{\textit{School of Artificial Intelligence and Computer Science} \\
\textit{Jiangnan University}\\
Wuxi, China \\
xiaojun\_wu\_jnu@163.com}
}

\maketitle

\begin{abstract}
In image fusion tasks, an ideal image decomposition method can bring better performance.
MDLatLRR has done a great job in this aspect, but there is still exist some space for improvement.
Considering that MDLatLRR focuses solely on the detailed parts (salient features) extracted from input images via latent low-rank representation (LatLRR), the basic parts (principal features) extracted by LatLRR are not fully utilized.
Therefore, we introduced an enhanced multi-level decomposition method named dual-decomposed MDLatLRR (D2-LRR) which effectively analyzes and utilizes all image features extracted through LatLRR.
Specifically, color images are converted into YUV color space and grayscale images, and the Y-channel and grayscale images are input into the trained parameters of LatLRR to obtain the detailed parts containing  four rounds of decomposition and the basic parts.
Subsequently, the basic parts are fused using an average strategy, while the detail part is fused using kernel norm operation.
The fused image is ultimately transformed back into an RGB image, resulting in the final fusion output. We apply D2-LRR to medical image fusion tasks. The detailed parts are fused employing a nuclear-norm operation, while the basic parts are fused using an average strategy. Comparative analyses among existing methods showcase that our proposed approach attains cutting-edge fusion performance in both objective and subjective assessments.
\end{abstract}


\section{Introduction}
Image fusion is a critical application within the realm of medical imaging. 
The primary focus of medical image fusion lies in the integration of modalities such as single-photon emission computed tomography (SPECT), magnetic resonance imaging (MRI), computerized tomography (CT), and positron emission tomography (PET) modalities.
By combining the high-resolution capabilities of CT scans for bones with MRI scans for soft tissue, the primary data is integrated to offer valuable assistance in clinical diagnosis, treatment planning, and ongoing observation. 
SPECT and PET can provide insights into the functional and metabolic aspects of the human body, albeit with limited spatial resolution.
The fusion of SPECT or PET with MRI images effectively reveals metabolic activity in the thalamus, nucleus, and cortex, along with alterations in blood circulation.

Within the realm of image fusion, the central challenge revolves around extracting characteristics from the input images and applying an adaptive approach to generate the fused images.

It is common knowledge that one of the most typical fusion methods in the transform domain is the multi-scale transform (MST).
Fusion methods based on MST typically involve three fundamental stages: decomposition, fusion, and reconstruction.
Many MST methods are been proposed, such as wavelet transform \cite{2}, pyramid \cite{3} , curvelet \cite{4}, contourlet \cite{5}, and shearlet \cite{6} etc.

Representation learning is also widely used \cite{lrrnet}. 
For medical image fusion, sparse representation (SR) \cite{8} combines nonsubsampled contourlet transform.
Furthermore, Low Rank Representation (LRR) has been used in image fusion\cite{9}.
The presented approach introduces an innovative multi-focus image fusion technique that relies on dictionary learning and LRR, achieving improved performance in terms of both global and local structures.

The progress in deep learning has given rise to a multitude of fusion methods that leverage deep learning principles.
In 2017, the convolutional neural network was employed to get fusion results by creating a weight map \cite{LiuY2017}.
In 2019, Li et al. \cite{DenseFuse} introduced an innovative fusion framework utilizing deep learning techniques for the integration of infrared and visible images.
In 2022, Zhao et al. \cite{CDDFuse} combined CNN and Transformer to decompose and fuse image features, achieving certain fusion results, but changing color information of source images.
In 2023, Cheng et al. \cite{MUFusion} adopted a deep learning method with memory units to generate fused images.
While deep learning-based techniques yield effective fusion results, they tend to give less consideration to the process of image decomposition.
Hence, a fresh approach to multi-level image decomposition known as MDLatLRR was proposed \cite{MDLatLRR}, which is based on LatLRR principles.

However, in the MDLatLRR approach, the authors exclusively employed the detail components obtained through LatLRR, omitting the base components. 
Hence, this paper introduces an enhanced version of MDLatLRR, referred to as D2-LRR, that maximizes the utilization of features acquired through LatLRR.

The primary contributions can be outlined as follows:

(1) D2-LRR explores features derived from the decomposition based on the latent low-rank representation and comprehensively utilizes them.

(2) The fusion outcome is meticulously analyzed to assess the impact of sparse noise.

(3) D2-LRR is utilized for medical image fusion and achieves improved fusion results both objectively and subjectively.

In this paper, the organization of the subsequent sections is as follows: Concise overviews of latent low-rank representation and MDLatLRR are provided in Section \ref{relatedworks}. A detailed presentation of D2-LRR and the fusion method are proposed in Section \ref{method}.
Section \ref{experiment} covers the experimental setup, results, and evaluations. Section \ref{conclusion} concludes the paper.

\section{Related Work}
\label{relatedworks}

\subsection{LatLRR: Latent Low-Rank Representation}
Latent Low-Rank Representation (LatLRR)\cite{LatLRR} was introduced in 2011 for the purpose of feature extraction and subspace segmentation. LatLRR is formulated as the subsequent optimization challenge:
\begin{align} \label {equ:lrr}
  \min_{Z,L,E}||Z||_* + ||L||_* + \lambda||E||_1 \\
  s.t., X = XZ + LX + E \nonumber 
\end{align}
In this context, $X$ stands for the observed matrix, $Z$ represents the low-rank coefficient matrix, $L$ signifies the projection matrix, $E$ denotes the sparse noise matrix, and $||\cdot||_*$ refers to the nuclear norm, while $||\cdot||_1$ represents the $\ell_1$-norm.
We refer to $E$ as the error term.
The parameter $\lambda >0$ is a parameter.

In the context of visual content \cite{LatLRR}, the matrix product $XZ$ encapsulates principal features and base information, whereas $LX$ emphasizes salient features and detailed information.

For our methodology, dividing the input image into $M$ patches is our assumption, each patch having a size of $n\times n$. Consequently, the matrix $X$ with $(n\times n)\times M$, the matrix $L$ with $(n\times n)\times(n\times n)$, and the matrix $Z$ with $M\times M$. It is not difficult to understand from the perspective of matrix dot product that the size of $L$ is associated with the image patch dimensions, allowing it to extract salient features from arbitrary-sized input images.
However, the size of $Z$ is related to the size of the input image.
In addition, this approach also brings a problem, the noise $E$ is difficult to separate.

\subsection{MDLatLRR}
\begin{figure}[!ht]
\centering
\includegraphics[width=0.8\linewidth]{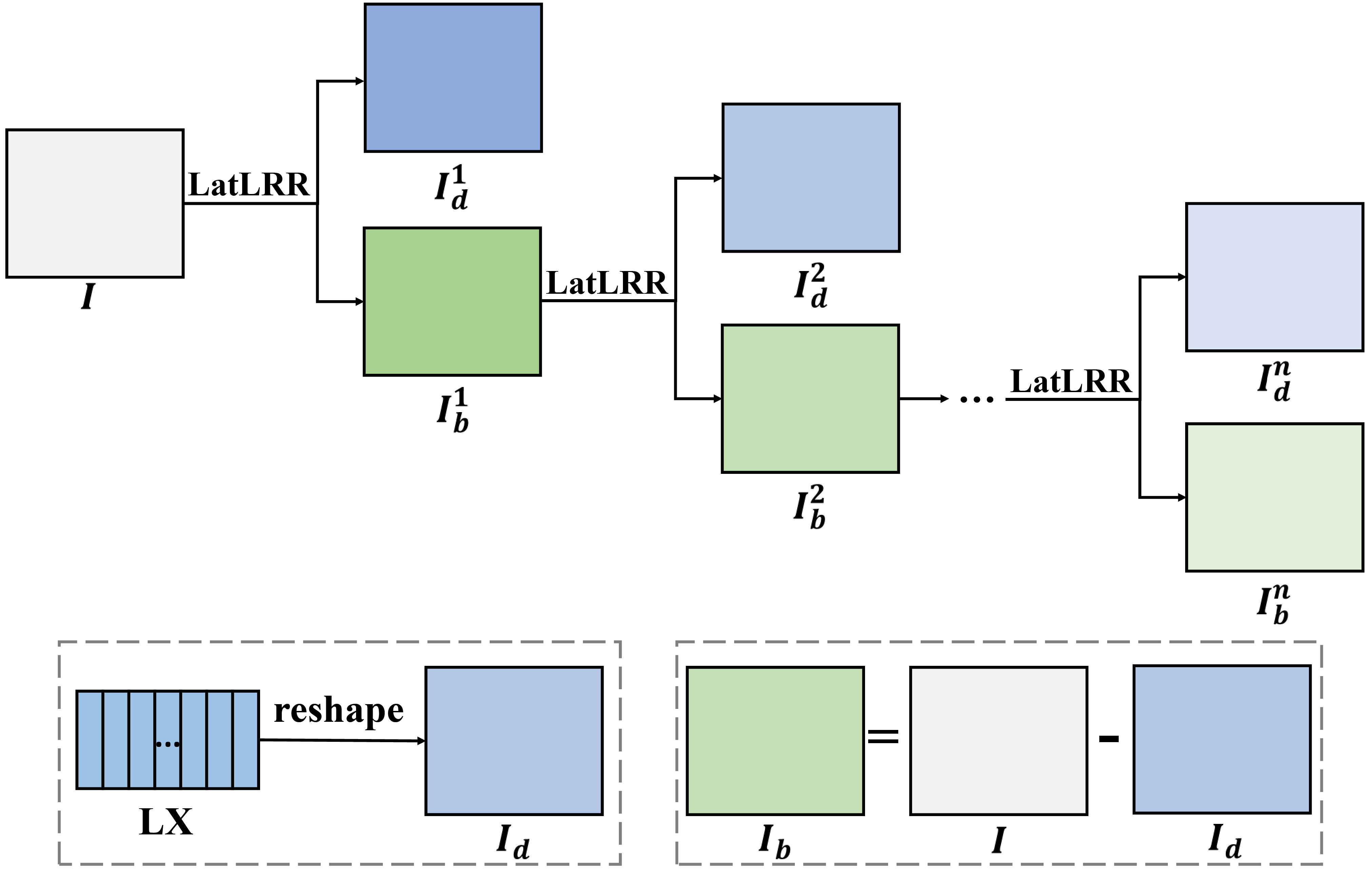}
\caption{MDLatLRR architecture.}
\label{fig:DLatLRR}
\end{figure}

Utilizing the principles of LatLRR, a groundbreaking multi-level image decomposition technique, MDLatLRR \cite{MDLatLRR}, was proposed.
This work brilliantly applies LatLRR to image fusion tasks.
In Fig.~\ref{fig:DLatLRR}, the schematic representation of MDLatLRR is visually represented.

In Fig.~\ref{fig:DLatLRR}, The input image ($I$) is decomposed into base parts ($I_b$) and detailed parts ($I_d$). Detailed parts are extracted using the projection matrix $L$, while base parts are acquired by subtracting detailed parts from the input image. This decomposition process is iteratively applied to the base parts. Consequently, after $n$ iterations, the result includes $n$ detailed parts and one final base part derived from the input image.Among it, the acquisition of the basic part is too simple and does not make full use of the information that has been decomposed.
The reason is that the projection matrix $Z$ is closely related to the size of the training image, while the image size in the test phase is not fixed.
The core improvement of this paper is precisely about this point.

\section{Methodology}
\label{method}

\subsection{$P$: A Low-Rank Projection Matrix}

In accordance with \eqref{equ:lrr}, due to the close correlation between $Z$ and the image size, it is not advisable to perform pre-training beforehand. Training for each individual image would inevitably result in a sharp decrease in the algorithm's execution speed. 
Therefore, in order to extract principal features from LatLRR, it is necessary to learn a low-rank projection matrix, denoted as $P$, which captures the principal features of the images. 
Motivated by sparse representation, the low-rank projection matrix $P$ is defined by:
\begin{align} \label {equ:low_rank_recovery}
\min\limits_{P}||P||_* \ s.t. XZ = PX
\end{align}

In equation \eqref{equ:low_rank_recovery}, the recovery result $PX$ represents base parts (principal features) of $X$, akin to $XZ$ in LatLRR. Moreover, $P$ is associated with the image patch size, similar to the role of $L$. Consequently, we employ $P$ to extract base parts, or principal features, from $X$.

In the end, both $L$ and $P$ are utilized to extract detailed and base parts of the input image. In contrast to MDLatLRR, our approach maximizes the utilization of the salient and principal features acquired through LatLRR. Thus, our method is termed D2-LRR, and it has been applied to medical image fusion techniques.

Furthermore, we conduct a detailed discussion on the incorporation of the "error" term ($E$) for fusion. Notably, our experiments reveal that the fusion performance improves when excluding the error term. In Section\ref{ablation}, we will delve into the specific experimental results in detail.

\subsection{Dual-decomposed MDLatLRR}
\begin{figure}[!ht]
\centering
\includegraphics[width=0.5\linewidth]{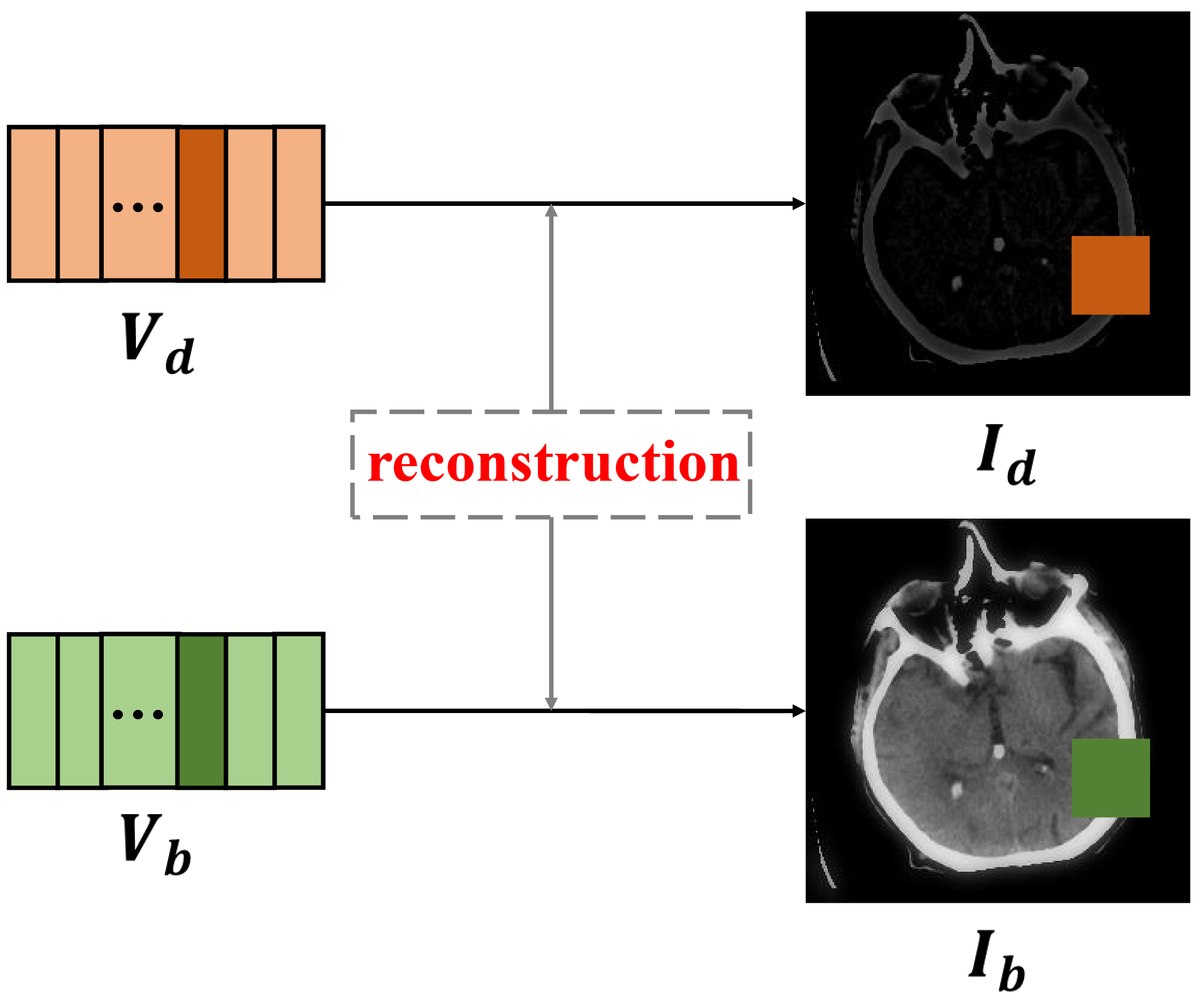}
\caption{The reconstruction graph.}
\label{fig:v2Recons}
\end{figure}

We partition the source image ($I$) into $M$ image patches using a sliding window approach with the $n\times n$ window and the step is 1.
These patches are then reconstructed into a matrix $X$ with $(n\times n)\times M$ dimension, where each column represents an image patch.
The basic and detailed data of $X$ are calculated in the following manner:
\begin{align} \label {equ:LX}
  V_b = PX\\
  V_d = LX
\end{align}
where $P$ and $L$ are acquired through learning using \eqref{equ:low_rank_recovery} and \eqref{equ:lrr}, respectively.
Here, $V_d$ represents detail part vectors and $V_b$ represents base part vectors. Utilizing $V_d$ and $V_b$, we reconstruct detail and base images, as demonstrated in Fig.~\ref{fig:v2Recons}. 
The reconstruction operator is used to transform each column of $V_d$ and $V_b$ into an image patch.
Pixels that overlap are handled by applying an averaging technique, which produces both the base image ($I_b$) and the detail image ($I_d$).
From a subjective point of view, it can be intuitively seen that $I_b$ as the basic part obviously contains more high-frequency information and texture information. 
The $I_d$ details section contains more outline information and detailed information.

\begin{figure}[!ht]
\centering
\includegraphics[width=\linewidth]{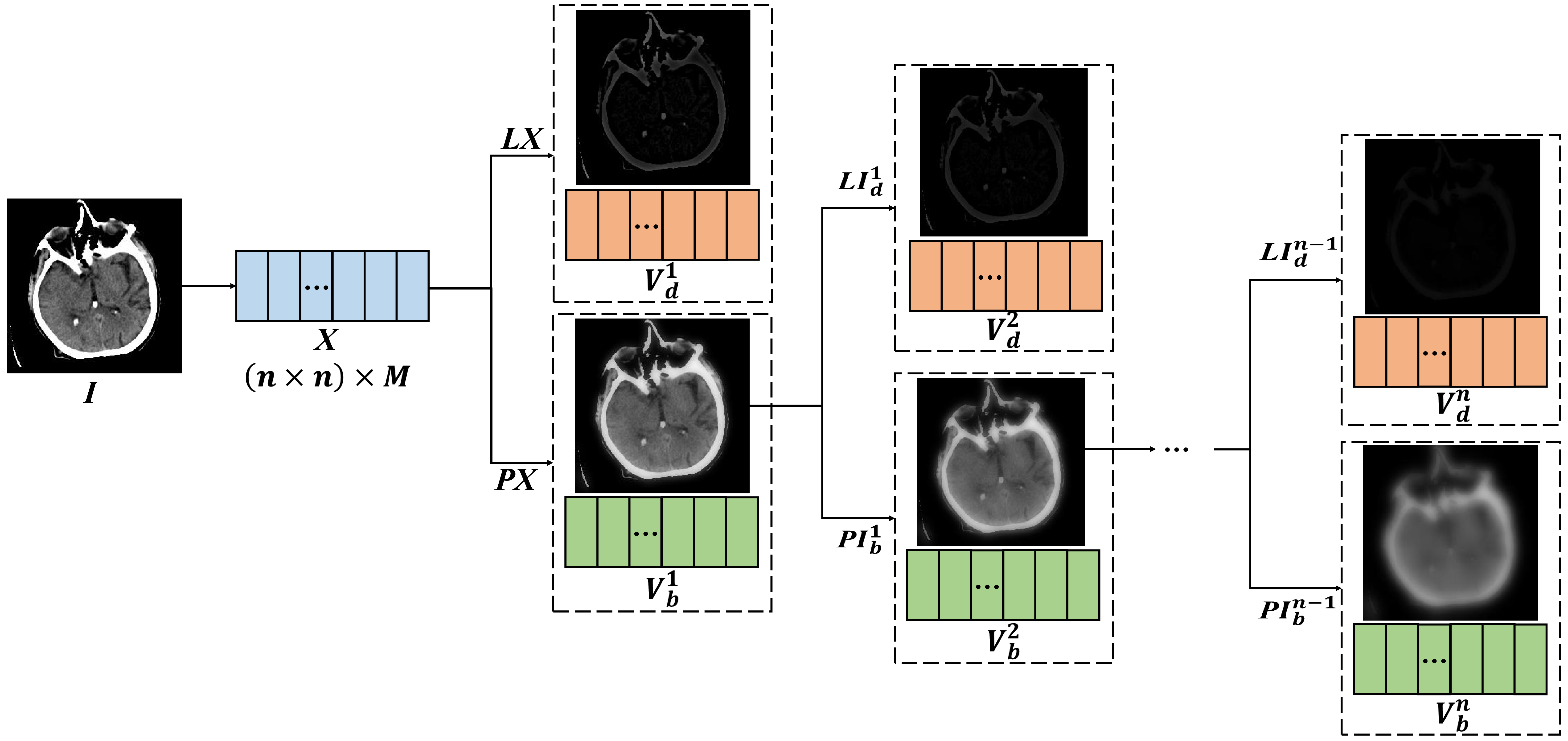}
\caption{D2-LRR framework.}
\label{fig:MDv2}
\end{figure}

Building upon the decomposition framework involving $L$ and $P$, basic parts are broken down several times. Consequently, in Fig.~\ref{fig:MDv2}, the architecture of the dual-decomposed MDLatLRR (D2-LRR) is depicted.

In Fig.~\ref{fig:MDv2}, $I$ means the input image, $n$ signifies the decomposition level, and $V_b^i$ and $V_d^i$ represent the base and detail part vectors of the $i$-th layer. 
Hence, through decomposition repeated $n$ times, $V_d^{1:n}$ and $V_b^n$ can be acquired.

\subsection{Acquiring knowledge of the Projection Matrices: $L$ and $P$}
\label{LP}
As the medical images for fusion are categorized into four types: CT, MRI, SPECT, and PET, ten images were chosen from each category for training purposes, amounting to a total of 40 training images.

The training approach corresponds to that of MDLatLRR. 
In order to have enough training samples in the training phase, we divided the existing samples into chunks.
We apply sliding window technology with the $16\times16$ window and the step is 16 to process the training dataset, resulting in 5955 image patches.
Subsequently, these patches are categorized into two groups: smooth image patches and detail image patches, determined by their Standard Deviation (SD) \cite{SDpaper} using equations \eqref{equ:SD} and \eqref{equ:chooseSD}.

\begin{align} \label {equ:SD}
  SD(patch)=\sqrt{\sum^H_{i=1}\sum^W_{j=1}(patch(i,j)-\mu)^2}
\end{align}

\begin{align}\label{equ:chooseSD}
  	C(patch)=\left\{\begin{array}{ll}
		detail & \textrm{ SD(patch)$>$Th }  \\
     smooth & \textrm{ otherwise}
	\end{array}\right.
\end{align}
Here, $patch$ represents the image patch and the threshold value Th is utilized to verify whether an image patch falls into the smooth or detail category. In this paper, Th is set to 0.5.

After classification, our training dataset is divided into 4292 detail image patches and 1663 smooth image patches. Randomly select a certain number of batches as the training set.
Subsequently, this finalized trained dataset is employed. It is aim to train the matrices $L$ and $P$ using equations \eqref{equ:lrr} and \eqref{equ:low_rank_recovery}.
After training is completed, $L$ and $P$ can be directly used in the testing phase.

\subsection{The Fusion Method for Medical Images Using D2-LRR}
\begin{figure}[!ht]
\centering
\includegraphics[width=\linewidth]{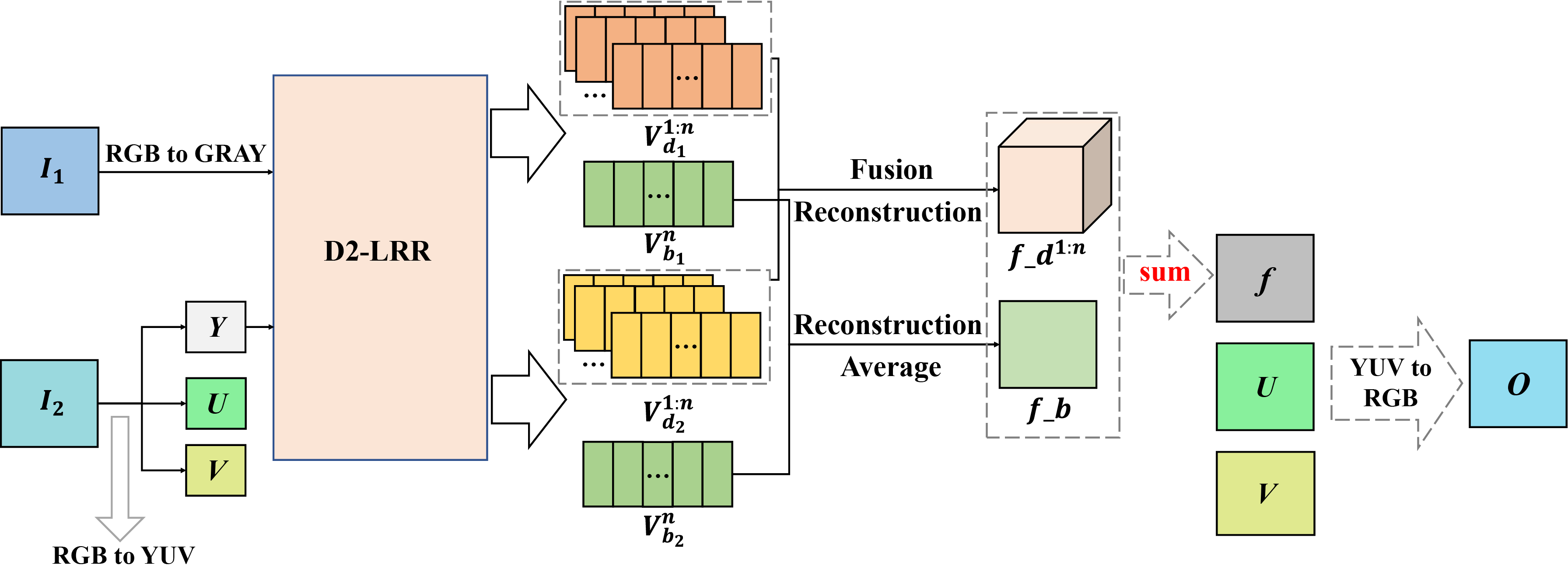}
\caption{The proposed fusion method framework.}
\label{fig:framework}
\end{figure}

After achieveing matrices $L$ and $P$ which has been trained, we introduce an innovative approach based on D2-LRR for fusion. 
General framework of this paper is illustrated in Fig.~\ref{fig:framework}.

\begin{table*}[!ht]
\centering
\caption{\label{tab:table1}The mean metrics scores for 20 fused images. $D2-LRR\_B\_D\_E\_level\_i$ denotes decomposition $i$ times based on $D2-LRR\_B\_D\_E$, and $D2-LRR\_B\_D\_level\_i$ denotes decomposition $i$ times based on $D2-LRR\_B\_D$. The highest values are highlighted in bold.}
\resizebox{\textwidth}{!}{
\begin{tabular}{|c|c|c|c|c|c|c|c|c|c|c|}
\hline
Images & Methods & Qabf & MI & FMI\_dct & FMI\_pixel & FMI\_w & $PSNR_a$ & EN & AG & $SSIM_a$ \\
\hline
\multirow{8}*{CT\_MRI}& D2-LRR\_B\_D\_E\_level\_1 & 0.563 & 8.683 & 0.406 & 0.868 & 0.449 & 15.495 & 4.341 & 0.118 & \textbf{0.753} \\
\cline{2-11}
&D2-LRR\_B\_D\_E\_level\_2 & 0.508 & 8.013 & 0.341 & 0.859 & 0.415 & 14.183 & 4.007 & 0.164 & 0.713 \\
\cline{2-11}
&D2-LRR\_B\_D\_E\_level\_3 & 0.388 & 6.995 & 0.283 & 0.835 & 0.392 & 12.832 & 3.498 & 0.191 & 0.674 \\
\cline{2-11}
&D2-LRR\_B\_D\_E\_level\_4 & 0.334 & 6.226  & 0.248 & 0.813 & 0.375 & 11.998 & 3.113 & \textbf{0.210} & 0.649 \\
\cline{2-11}
&D2-LRR\_B\_D\_level\_1 & 0.415 & \textbf{8.855} & \textbf{0.424} & 0.869 & \textbf{0.461} & \textbf{15.722} & \textbf{4.427} & 0.083 & 0.752 \\
\cline{2-11}
&D2-LRR\_B\_D\_level\_2 & \textbf{0.564} & 8.666 & 0.404 & \textbf{0.869} & 0.445 & 15.459 & 4.333 & 0.121 & 0.751 \\
\cline{2-11}
&D2-LRR\_B\_D\_level\_3 & 0.546 & 8.266 & 0.365 & 0.865 & 0.424 & 14.675 & 4.133 & 0.152 & 0.729 \\
\cline{2-11}
&D2-LRR\_B\_D\_level\_4 & 0.463 & 7.591 & 0.322 & 0.854 & 0.405 & 13.707 & 3.795 & 0.171 & 0.702 \\
\hline
\multirow{8}*{MRI\_PET} & D2-LRR\_B\_D\_E\_level\_1 & 6.642 & 6.917 & 0.408 & 0.873 & 0.447 & 19.505 & 3.458 & 0.099 & \textbf{0.803} \\
\cline{2-11}
&D2-LRR\_B\_D\_E\_level\_2 & 0.517 & 6.472 & 0.321 & 0.855 & 0.407 & 16.690 & 3.236 & 0.153 & 0.770 \\
\cline{2-11}
&D2-LRR\_B\_D\_E\_level\_3 & 0.362 & 5.961 & 0.273 & 0.835 & 0.380 & 14.504 & 2.980 & 0.188 & 0.732 \\
\cline{2-11}
&D2-LRR\_B\_D\_E\_level\_4 & 0.289 & 5.477 & 0.247 & 0.821 & 0.367 & 13.164 & 2.739 & \textbf{0.212} & 0.708 \\
\cline{2-11}
&D2-LRR\_B\_D\_level\_1 & 0.560 & \textbf{7.102} & \textbf{0.448} & \textbf{0.880} & \textbf{0.469} & \textbf{19.828} & \textbf{3.551} & 0.067 & 0.800 \\
\cline{2-11}
&D2-LRR\_B\_D\_level\_2 & \textbf{0.644} & 6.895 & 0.399 & \textbf{0.872} & 0.443 & 19.410 & 3.448 & 0.104 & 0.802 \\
\cline{2-11}
&D2-LRR\_B\_D\_level\_3 & 0.577 & 6.622 & 0.342 & 0.862 & 0.418 & 17.549 & 3.311 & 0.139 & 0.783 \\
\cline{2-11}
&D2-LRR\_B\_D\_level\_4 & 0.464 & 6.313 & 0.302 & 0.849 & 0.397 & 15.823 & 3.156 & 0.164 & 0.758 \\
\hline
\multirow{8}*{MRI\_SPECT} & D2-LRR\_B\_D\_E\_level\_1 & 6.686 & 8.497 & 0.375 & 0.865 & 0.406 & 19.728 & 4.248 & 0.077 & \textbf{0.755} \\
\cline{2-11}
&D2-LRR\_B\_D\_E\_level\_2 & 0.479 & 8.333 & 0.306 & 0.846 & 0.371 & 16.619 & 4.167 & 0.135 & 0.704 \\
\cline{2-11}
&D2-LRR\_B\_D\_E\_level\_3 & 0.289 & 7.788 & 0.254 & 0.821 & 0.344 & 13.710 & 3.894 & 0.178 & 0.652 \\
\cline{2-11}
&D2-LRR\_B\_D\_E\_level\_4 & 0.222 & 7.151 & 0.224 & 0.799 & 0.323 & 11.987 & 3.576 & \textbf{0.207} & 0.620 \\
\cline{2-11}
&D2-LRR\_B\_D\_level\_1 & 0.611 & \textbf{8.594} & \textbf{0.417} & \textbf{0.875} & \textbf{0.433} & \textbf{19.987} & \textbf{4.297} & 0.051 & 0.752 \\
\cline{2-11}
&D2-LRR\_B\_D\_level\_2 & \textbf{0.686} & 8.497 & 0.367 & 0.864 & 0.401 & 19.663 & 4.249 & 0.082 & 0.753 \\
\cline{2-11}
&D2-LRR\_B\_D\_level\_3 & 0.574 & 8.443 & 0.326 & 0.853 & 0.384 & 17.771 & 4.221 & 0.118 & 0.723 \\
\cline{2-11}
&D2-LRR\_B\_D\_level\_4 & 0.406 & 8.214 & 0.286 & 0.839 & 0.363 & 15.520 & 4.107 & 0.149 & 0.688 \\
\hline
\end{tabular}}
\end{table*}

In Fig.~\ref{fig:framework}, two sets of features $V_{d_i}^{1:n}$ and $V_{b_i}^n$ are obtained, where $n$ signifies the number of decompositions and $i\in{1,\dots,k}$, $k$ represents the number of input images.
Two sets of features have been acquired, $V_{d_i}^{1:n}$ and $V_{b_i}^n$. 
Here, $n$ represents the number of decompositions, while $i\in{1,\dots,k}$, $k$ denotes the count of input images.
We set $k=2$ and $n\in{1,2,3,4}$. The features $V_{d_i}^{1:n}$ are fused using a nuclear-norm-based fusion strategy\cite{MDLatLRR}, as outlined in \eqref{equ:norm1}. Subsequently, the detail image is obtained through the reconstruction operator. 
The features $V_{b_i}^n$ reconstruct the base image, which is then fused using an averaging strategy.
\begin{align} \label {equ:norm1}
V_{d_f}^j=\sum^k_{i=1}W_i^j\times V_{d_i}^j\\
W_i^j=\frac{w_i^j}{\sum_{i=1}^kw_i^j}\\
w_i^j = ||re(V_{d_i}^j)||_*
\end{align}
Here, $re(\cdot)$ represents the operator utilized for reconstructing image patches from $V_{d_i}^j$, where $j$ corresponds to the column of $V_{d_i}$. Additionally, $||\cdot||_*$ stands for the nuclear norm, which accumulate all singular values of the matrix.
Once the fused basic images ($f_b$) obtained and detailed images ($f_d^{1:n}$), we calculate the fusion results using \eqref{equ:fusion}.
This paper performs an accumulation operation on them to obtain fused images.

\begin{align} \label {equ:fusion}
f=f\_b + \sum_{i=1}^nf\_d^i
\end{align}

The source images, including those that appear black and white, are all 3-channel color images.
This paper employ the technique outlined in \cite{YinM2018} to convert color images from RGB channels into the YUV color space.
Next, we employ the proposed fusion technique to merge the grayscale image and the Y channels. Specifically, in the case of the CT and MRI image set, the CT image is converted to grayscale and then fused with the Y channels of the MRI image. 
Likewise, for the MRI and PET image set, the MRI image is transformed into grayscale and combined with the Y channels of the PET image using the same approach.
The same procedure is applied to the MRI and SPECT image set.

\section{Experiments and Assessment}
\label{experiment}
\subsection{Experiment Configuration}
\begin{figure}[!ht]
\centering
\includegraphics[width=0.6\linewidth]{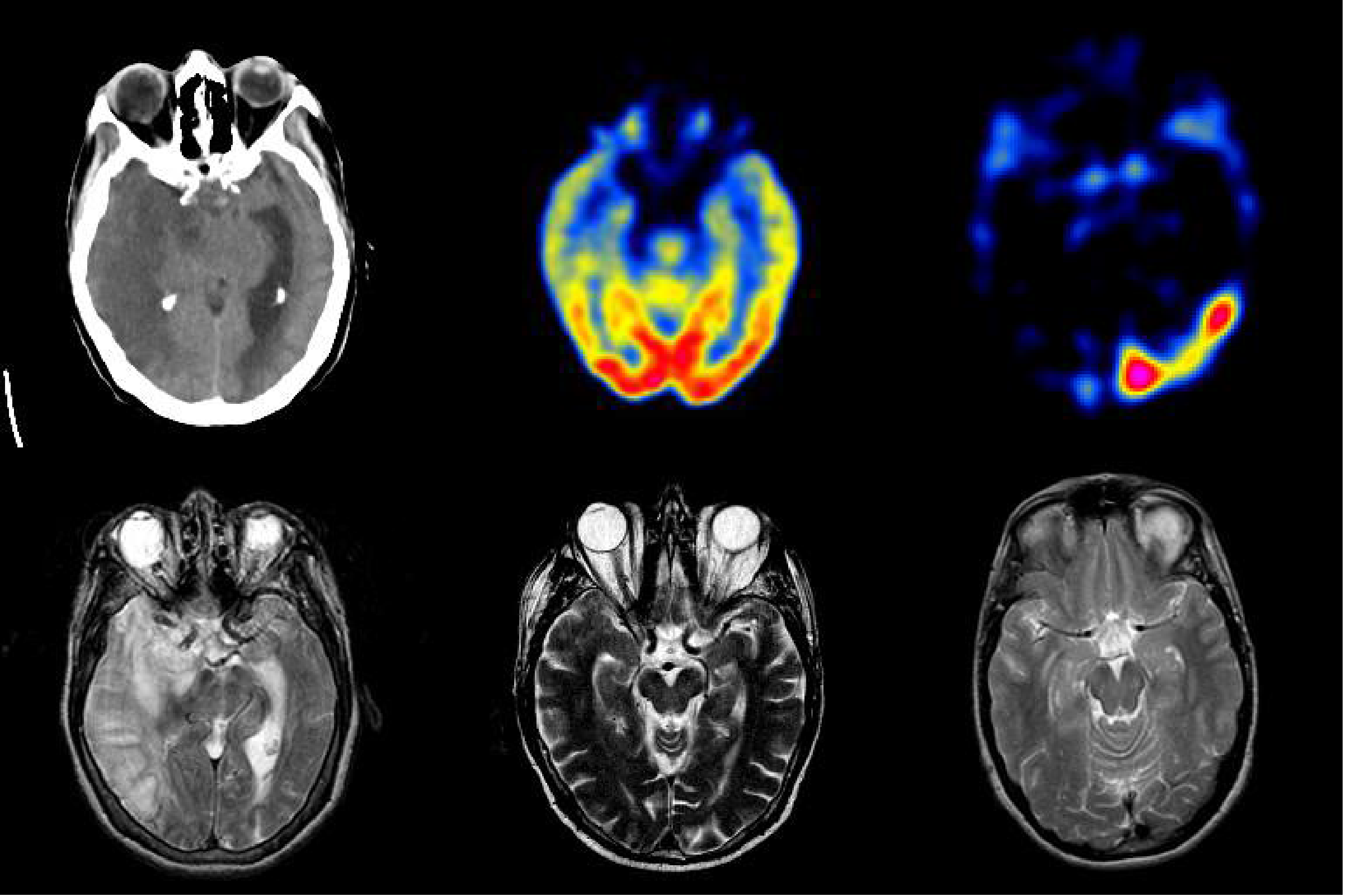}
\caption{There are three sets: the first row comprises CT, PET, and SPECT, while the second consists of MRI images.}
\label{fig:HVMD}
\end{figure}

During this experimentation, we have three distinct fusion scenarios, namely, the fusion of MRI with CT, MRI with SPECT and MRI with PET \cite{data}, as illustrated in Fig.~\ref{fig:HVMD}.

\begin{table*}[!ht]
\centering
\caption{\label{tab:table2}The mean metrics scores for 20 fused images. $MDLatLRR\_level\_i$ denotes decomposition $i$ times based on $MDLatLRR$, and $D2-LRR\_level\_i$ denotes decomposition $i$ times based on $D2-LRR$. The highest values are highlighted in bold.}
\resizebox{\textwidth}{!}{
\begin{tabular}{|c|c|c|c|c|c|c|c|c|c|c|}
\hline
Images & Methods & Qabf & MI & $PSNR_a$ & EN & AG & FMI\_pixel & $SSIM_a$ & MS\_SSIM & SCD \\
\hline
\multirow{8}*{CT\_MRI} & MDLatLRR\_level\_1 & 0.416 & 8.853 & 15.722 & 4.427 & 0.084 & 0.869 & \textbf{0.753} & 0.882 & 1.036 \\
\cline{2-11}
&MDLatLRR\_level\_2 & \textbf{0.565} & 8.664 & 15.457 & 4.332 & 0.122 & \textbf{0.870} & 0.752 & 0.945 & 1.161 \\
\cline{2-11}
&MDLatLRR\_level\_3 & 0.546 & 8.261 & 14.672 & 4.131 & 0.153 & 0.866 & 0.729 & 0.949 & 1.257 \\
\cline{2-11}
&MDLatLRR\_level\_4 & 0.463 & 7.586 & 13.705 & 3.793 & \textbf{0.172} & 0.854 & 0.702 & 0.920 & 1.290 \\
\cline{2-11}
&D2-LRR\_level\_1 & 0.415 & \textbf{8.855} & \textbf{15.723} & \textbf{4.428} & 0.083 & 0.869 & 0.752 & 0.882 & 1.035 \\
\cline{2-11}
&D2-LRR\_level\_2 & 0.564 & 8.666 & 15.459 & 4.333 & 0.121 & 0.869 & 0.751 & 0.945 & 1.161 \\
\cline{2-11}
&D2-LRR\_level\_3 & 0.546 & 8.266 & 14.675 & 4.133 & 0.152 & 0.865 & 0.729 & \textbf{0.950} & 1.257 \\
\cline{2-11}
&D2-LRR\_level\_4 & 0.463 & 7.591 & 13.707 & 3.795 & 0.171 & 0.854 & 0.702 & 0.921 & \textbf{1.291} \\
\hline
\multirow{8}*{MRI\_PET} & MDLatLRR\_level\_1 & 0.561 & 7.010 & \textbf{19.829} & 3.550 & 0.067 & 0.879 & 0.710 & 0.937 & 1.101 \\
\cline{2-11}
&MDLatLRR\_level\_2 & \textbf{0.645} & 6.894 & 19.409 & 3.447 & 0.106 & 0.872 & \textbf{0.803} & 0.965 & 1.365 \\
\cline{2-11}
&MDLatLRR\_level\_3 & 0.577 & 6.619 & 17.545 & 3.309 & 0.140 & 0.862 & 0.783 & 0.941 & 1.410 \\
\cline{2-11}
&MDLatLRR\_level\_4 & 0.464 & 6.309 & 15.819 & 3.155 & \textbf{0.166} & 0.849 & 0.758 & 0.906 & 1.400 \\
\cline{2-11}
&D2-LRR\_level\_1 & 0.560 & \textbf{7.102} & 19.828 & \textbf{3.551} & 0.067 & \textbf{0.880} & 0.800 & 0.937 & 1.100 \\
\cline{2-11}
&D2-LRR\_level\_2 & 0.644 & 6.895 & 19.410 & 3.449 & 0.104 & 0.872 & 0.802 & \textbf{0.966} & 1.364 \\
\cline{2-11}
&D2-LRR\_level\_3 & 0.577 & 6.622 & 17.549 & 3.311 & 0.139 & 0.862 & 0.783 & 0.941 & \textbf{1.411} \\
\cline{2-11}
&D2-LRR\_level\_4 & 0.464 & 6.313 & 15.823 & 3.156 & 0.164 & 0.849 & 0.758 & 0.906 & 1.402 \\
\hline
\multirow{8}*{MRI\_SPECT} & MDLatLRR\_level\_1 & 0.612 & \textbf{8.595} & 19.987 & \textbf{4.297} & 0.051 & 0.874 & 0.752 & 0.937 & 0.888 \\
\cline{2-11}
&MDLatLRR\_level\_2 & \textbf{0.687} & 8.497 & 19.661 & 4.249 & 0.082 & 0.864 & 0.753 & 0.968 & 1.013 \\
\cline{2-11}
&MDLatLRR\_level\_3 & 0.573 & 8.442 & 17.767 & 4.221 & 0.119 & 0.853 & 0.723 & 0.924 & 1.093 \\
\cline{2-11}
&MDLatLRR\_level\_4 & 0.405 & 8.212 & 15.516 & 4.106 & \textbf{0.150} & 0.839 & 0.687 & 0.861 & 1.116 \\
\cline{2-11}
&D2-LRR\_level\_1 & 0.611 & 8.594 & \textbf{19.988} & 4.297 & 0.051 & \textbf{0.875} & 0.752 & 0.937 & 0.888 \\
\cline{2-11}
&D2-LRR\_level\_2 & 0.686 & 8.497 & 19.663 & 4.249 & 0.082 & 0.864 & \textbf{0.754} & \textbf{0.969} & 1.013 \\
\cline{2-11}
&D2-LRR\_level\_3 & 0.574 & 8.443 & 17.771 & 4.221 & 0.118 & 0.853 & 0.724 & 0.924 & 1.093 \\
\cline{2-11}
&D2-LRR\_level\_4 & 0.406 & 8.214 & 15.520 & 4.107 & 0.149 & 0.839 & 0.688 & 0.861 & \textbf{1.117}\\
\hline
\end{tabular}}
\end{table*}

Next, this paper evaluates this method by comparing it with various fusion methods in recent years, including: Image fusion based on full Convolutional neural network (IFCNN) \cite{IFCNN}, Unified Fusion Network Based on Similarity Constraints (U2Fusion) \cite{U2Fusion}, Medical Image Fusion Network Using Unsupervised Enhancement (EMFusion) \cite{EMFusion}, Multimodal Image Fusion Network (MUFusion) with Memory Units \cite{MUFusion}. 
The concept of feature separation is incorporated into the multimodal image fusion networks of image fusion\cite{CDDFuse}, medical image fusion using sparse representation and three-level decomposition\cite{TL-SR}, and MDLatLRR\cite{MDLatLRR}, respectively.

To assess various methods, we have chosen specific metrics to gauge the quality of the images.
They are: average gradient (AG); visual information fidelity (VIF) \cite{VIF}; standard deviation (SD) \cite{SD}; Mutual Information (MI) \cite{MI}; entropy (EN); Qabf \cite{Qabf}; $FMI_{pixel}$, $FMI_{dct}$ and $FMI_w$ \cite{Fast-FMI}; a new no-reference image fusion performace measure (MS\_SSIM) \cite{MS_SSIM}; the sum of correlations of differences (SCD) \cite{SCD}; $PSNR_a$ and $SSIM_a$, computed using \eqref{equ:psnr} and \eqref{equ:SSIM}.
\begin{equation} 
PSNR_a(F)=(PSNR(F, I_1)+PSNR(F, I_2))\times0.5 \label{equ:psnr}
\end{equation}
\begin{align} 
SSIM_a(F)=(SSIM(F, I_1)+SSIM(F, I_2))\times0.5 \label{equ:SSIM}
\end{align}
Here, PSNR represents peak signal-to-noise ratio and SSIM represents structural similarity\cite{SSIM}.

Regarding the experimental configuration, 60 images were used for fusion, with 20 images in each group.
These images are obtained from the same source as cited in \cite{data}.
The fusion algorithms were executed using Matlab R2016a.

\subsection{Ablation Analysis}
\label{ablation}

We will assess how $E$ affects fusion performance.
Therefore, there are two sets of ablation experiments within the D2-LRR framework. 
The initial approach involves breaking down the image ($I$) into three components (detail part ($D$), base part ($B$), and error ($E$, $E=I-B-D$)) using the LatLRR method, referred to as D2-LRR\_B\_D\_E.
The second approach involves using only $B$ and $D$ components without including $E$, known as D2-LRR\_B\_D.
We employ two techniques for the fusion process and assess the fusion outcomes using specific metrics across three sets of medical image pairs (comprising 20 pairs in each group), as presented in Table \ref{tab:table1}.

It's evident that the fusion quality of D2-LRR\_B\_D surpasses that of D2-LRR\_B\_D\_E.
Due to the separation of sparse noise by LatLRR, the remaining features are devoid of artifacts and better suited for the fusion process.
Hence, we opt for D2-LRR\_B\_D as the ultimate iteration of our proposed technique. 
In the subsequent experiments, D2-LRR decomposes the image into a base component and a detail component, while disregarding the error (sparse noise).

\subsection{Baseline}
Since the approach builds upon the enhancements made to MDLatLRR, we conduct a comparison between D2-LRR and MDLatLRR using three sets of image pairs (with 20 pairs in each group).
In addition, certain indicators were selected for evaluating, as illustrated in Table \ref{tab:table2}.

\begin{figure}[!ht]
\centering
\includegraphics[width=\linewidth]{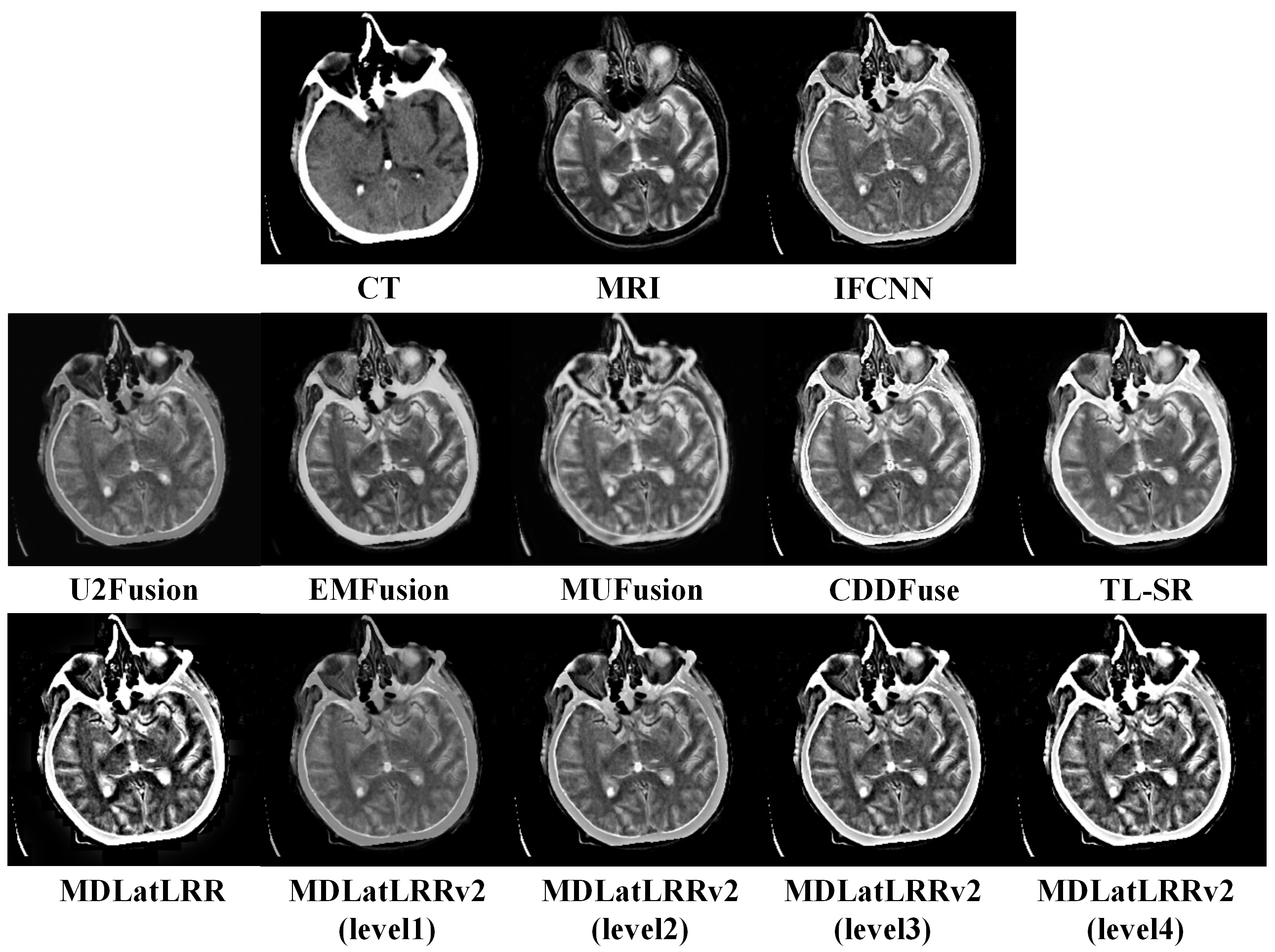}
\caption{
CT and MRI fusion results.}
\label{fig:ct_mr1}
\end{figure}
\begin{figure}[!ht]
\centering
\includegraphics[width=\linewidth]{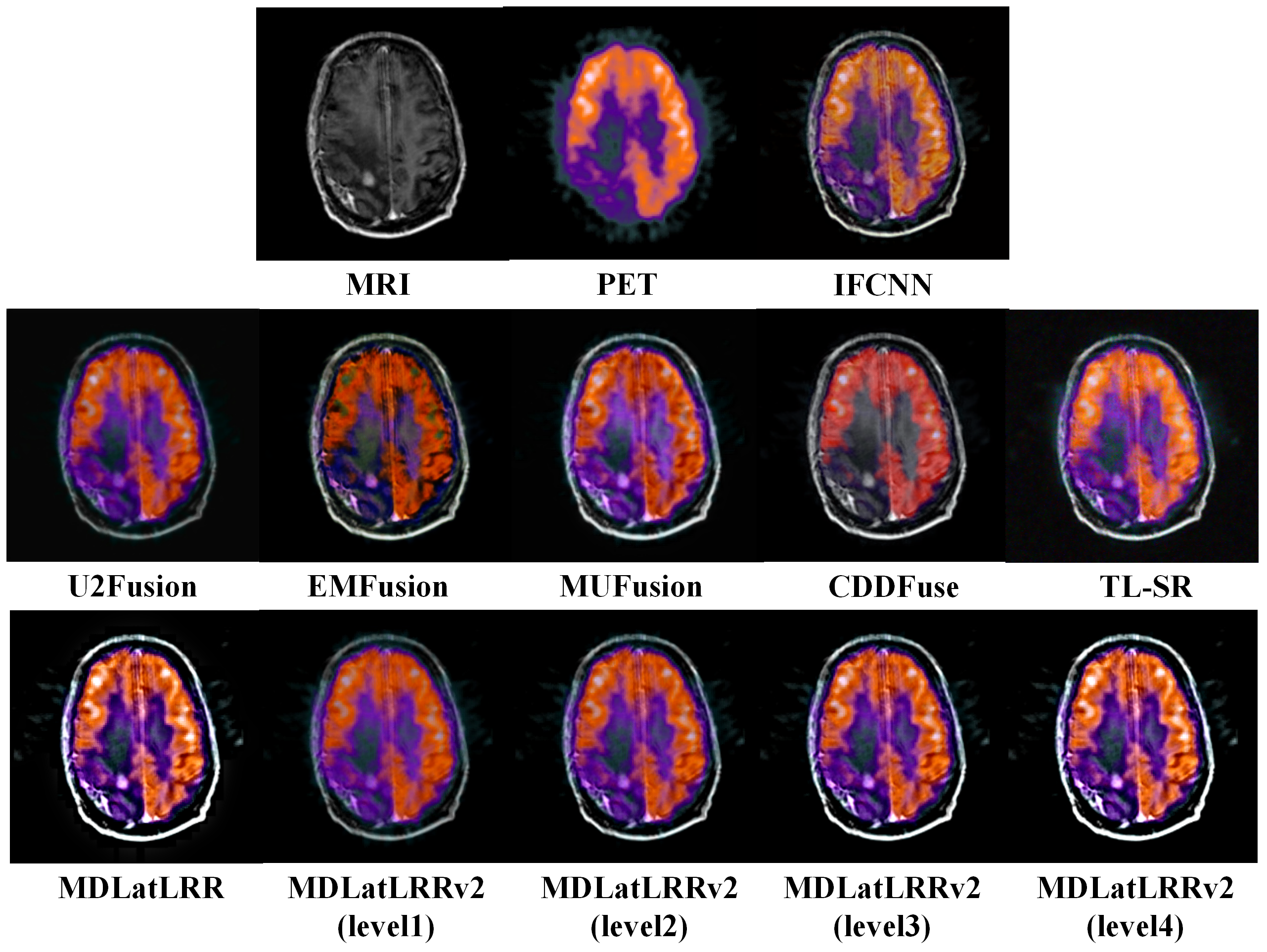}
\caption{
MRI and PET fusion results.}
\label{fig:mr_pet_1}
\end{figure}
\begin{figure}[!ht]
\centering
\includegraphics[width=\linewidth]{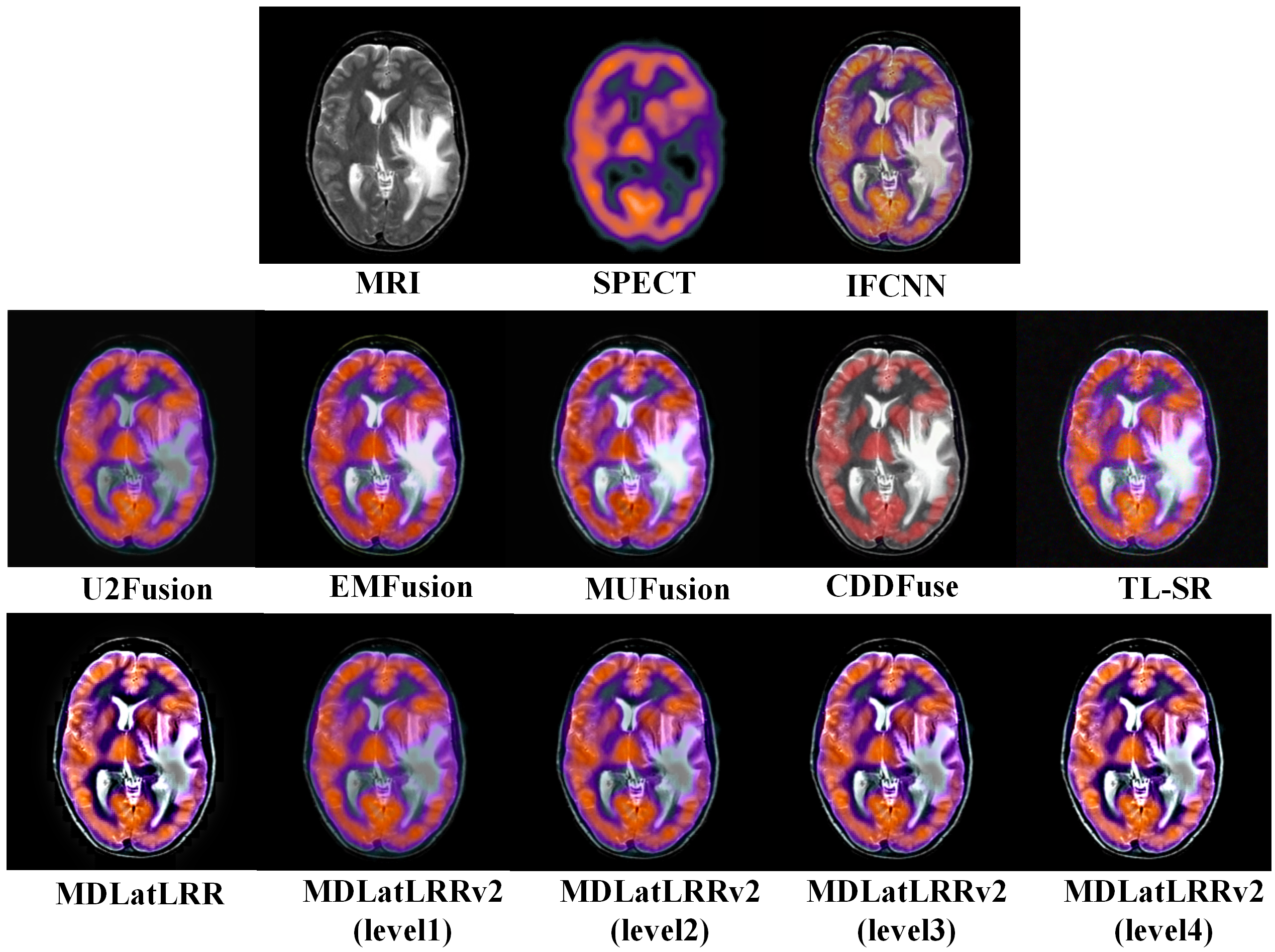}
\caption{
MRI and SPECT fusion results.}
\label{fig:mr_spect_1}
\end{figure}

It's evident that D2-LRR outperforms MDLatLRR in terms of numerical superiority across most of the evaluation criteria.
This demonstrates that D2-LRR more comprehensively and efficiently makes use of the image information acquired through LatLRR while eliminating sparse noise.
In addition, it can be seen from the indicators that the results of this paper have higher structural similarity than the source images and better retain the structural information of the source images.
Moving forward, we proceed with the comparison of the fusion technique using D2-LRR against other fusion methods.
We exclusively get the fusion results following four decompositions in MDLatLRR as the benchmark for comparison.

\subsection{Assessment from Subjective and Objective Perspectives}
This paper selects a pair of images from each series of experiments as examples and analyzes their visual effects as depicted in Fig.~\ref{fig:ct_mr1} - Fig.~\ref{fig:mr_spect_1}.

It is observable that in the case of CT and MRI fusion, IFCNN, U2Fusion, EMFusion, and MUFusion exhibit lower levels of CT information compared to the other methods. Moreover, in the two remaining experimental groups, U2Fusion retains less MRI information in the fusion results than the alternatives. 
Clearly, it is apparent that the fusion outcomes from EMFusion exhibit noise that impacts visual perception. 
Furthermore, both EMFusion and CDDFuse have altered original color in their results. 
The results of TL-SR, MDLatLRR and D2-LRR are more consistent with the perception of the human eye. 
Moreover, we assess all methods from an objective standpoint, and the findings are presented in Table \ref{tab:table7}-Table \ref{tab:table9}.

\begin{table}[!ht]
\centering
\caption{\label{tab:table7}The mean quality metrics values for 20 CT and MRI fused images.}
\resizebox{3.6in}{!}{
\begin{tabular}{|c|c|c|c|c|c|c|}
\hline
CT\_MRI & VIF & AG & SD & FMI\_dct & FMI\_w & MS\_SSIM \\
\hline
IFCNN & 0.6003 & 8.3779 & 76.6345 & 0.3779 & 0.4171 & 0.9424  \\
\hline
U2Fusion & 0.3148 & 4.6460 & 57.6950 & 0.3095 & 0.2420 & 0.8511 \\
\hline
EMFusion & 0.4782 & 5.6179 & 70.8714 & 0.3049 & 0.3483 & 0.8914 \\
\hline
MUFusion & 0.4897 & 6.4662 & 68.1673 & 0.3255 & 0.3111 & 0.8523 \\
\hline
CDDFuse & 0.6581 & 8.9587 & 83.1489 & 0.3394 & 0.3977 & 0.9286
\\
\hline
TL-SR & 0.6809 & 7.5238 & 86.4074  & {\color{blue}{\textbf{0.4112}}} & {\color{blue}{\textbf{0.4561}}} & 0.9402\\
\hline
MDLatLRR & {\color{red}{\textbf{1.0571}}} & {\color{red}{\textbf{10.5751}}} & {\color{blue}{\textbf{87.2797}}} & 0.3211 & 0.2989 & 0.8971 \\
\hline
D2-LRR\_level1 & 0.3506 & 5.0780 & 61.1988 & {\color{red}{\textbf{0.4243}}} & {\color{red}{\textbf{0.4611}}} & 0.8815 \\
\hline
D2-LRR\_level2 & 0.5695 & 7.2384 & 70.0686 & 0.4044 & 0.4454 & {\color{blue}{\textbf{0.9448}}} \\
\hline
D2-LRR\_level3 & 0.8379 & 9.2837 & 80.9316 & 0.3646 & 0.4242 & {\color{red}{\textbf{0.9490}}} \\
\hline
D2-LRR\_level4 & {\color{blue}{\textbf{1.0364}}} & {\color{blue}{\textbf{10.4513}}} & {\color{red}{\textbf{88.7297}}} & 0.3216 & 0.4047 & 0.9205 \\
\hline
\end{tabular}}
\end{table}

\begin{table}[!ht]
\centering
\caption{\label{tab:table8}The mean quality metrics values for 20 MRI and PET fused images.}
\resizebox{3.6in}{!}{
\begin{tabular}{|c|c|c|c|c|c|c|}
\hline
MRI\_PET & VIF & AG & SD & FMI\_dct & FMI\_w & MS\_SSIM \\
\hline
IFCNN & 0.8115 & 6.3813 & 55.1841 & 0.4023 & 0.4352 & {\color{red}{\textbf{0.9673}}}  \\
\hline
U2Fusion & 0.4437 & 3.3608 & 40.6198 & 0.1871 & 0.2420 & 0.8990 \\
\hline
EMFusion & 0.7124 & 5.8211 & 48.9050 & {\color{blue}{\textbf{0.4169}}} & 0.4160 & 0.9094 \\
\hline
MUFusion & 0.8025 & 5.5432 & 53.7949 & 0.3858 & 0.3363 & 0.9271 \\
\hline
CDDFuse & 0.8743 & 6.3213 & 58.6329 & 0.4197 & 0.3535 & 0.9500
\\
\hline
TL-SR & 0.7631 & 6.4090 & 57.1315  & 0.3009 & 0.1389 & 0.9342\\
\hline
MDLatLRR & {\color{red}{\textbf{1.6648}}} & {\color{blue}{\textbf{9.3301}}} & {\color{blue}{\textbf{68.9965}}} & 0.3033 & 0.2701 & 0.8976 \\
\hline
D2-LRR\_level1 & 0.5007 & 4.1290 & 44.6084 & {\color{red}{\textbf{0.4477}}} & {\color{red}{\textbf{0.4686}}} & 0.9373 \\
\hline
D2-LRR\_level2 & 0.8697 & 6.1459 & 52.7932 & 0.3992 & {\color{blue}{\textbf{0.4426}}} & {\color{blue}{\textbf{0.9648}}} \\
\hline
D2-LRR\_level3 & 1.3015 & 8.0470 & 62.2800 & 0.3423 & 0.4177 & 0.9411 \\
\hline
D2-LRR\_level4 & {\color{blue}{\textbf{1.6562}}} & {\color{red}{\textbf{9.3491}}} & {\color{red}{\textbf{69.7802}}} & 0.3015 & 0.3967 & 0.9059 \\
\hline
\end{tabular}}
\end{table}

\begin{table}[!ht]
\centering
\caption{\label{tab:table9}The mean quality metrics values for 20 MRI and SPECT fused images.}
\resizebox{3.6in}{!}{
\begin{tabular}{|c|c|c|c|c|c|c|}
\hline
MRI\_SPECT & VIF & AG & SD & FMI\_dct & FMI\_w & MS\_SSIM \\
\hline
IFCNN & 0.8816 & 6.1111 & 52.0373 & 0.3711 & 0.3987 & {\color{blue}{\textbf{0.9672}}}  \\
\hline
U2Fusion & 0.4705 & 3.0746 & 38.7540 & 0.2777 & 0.2158 & 0.8761 \\
\hline
EMFusion & 0.7051 & 5.0099 & 50.5929 & 0.3677 & 0.3714 & 0.9432 \\
\hline
MUFusion & 0.9272 & 5.4609 & 58.2797 & 0.3627 & 0.3321 & 0.9279 \\
\hline
CDDFuse & 0.9723 & 6.0223 & 60.2942 & {\color{red}{\textbf{0.4264}}} & {\color{blue}{\textbf{0.4040}}} & 0.9541
\\
\hline
TL-SR & 0.8140 & 5.8939 & 54.6259  & 0.2376 & 0.1531 & 0.9234\\
\hline
MDLatLRR & {\color{red}{\textbf{2.3711}}} & {\color{blue}{\textbf{10.5380}}} & {\color{blue}{\textbf{68.2742}}} & 0.2890 & 0.2810 & 0.8531 \\
\hline
D2-LRR\_level1 & 0.5756 & 3.9003 & 42.6308 & {\color{blue}{\textbf{0.4166}}} & {\color{red}{\textbf{0.4330}}} & 0.9365 \\
\hline
D2-LRR\_level2 & 1.0083 & 5.9933 & 49.2444 & 0.3673 & 0.4008 & {\color{red}{\textbf{0.9681}}} \\
\hline
D2-LRR\_level3 & 1.6816 & 8.5310 & 59.4604 & 0.3258 & 0.3837 & 0.9237 \\
\hline
D2-LRR\_level4 & {\color{blue}{\textbf{2.3331}}} & {\color{red}{\textbf{10.6550}}} & {\color{red}{\textbf{69.6247}}} & 0.2864 & 0.3625 & 0.8606 \\
\hline
\end{tabular}}
\end{table}

As evident from the three tables above, where red bold indicates the maximum values and blue highlights the second-highest, it is easy to find that our work have more number one objective metrics, which signify that our method retains a greater amount of prominent details from source images, ultimately improving visual perception.
The second-highest value (VIF, AG) suggests that the fusion outcomes exhibit reduced noise, resulting in a more natural and clear appearance.

\section{Conclusion}
\label{conclusion}
Improved decomposition framework has been proposed, named D2-LRR, built upon MDLatLRR. D2-LRR primarily leverages image features obtained by LatLRR, both from the base level and the detail level, effectively. 
Additionally, we incorporate D2-LRR into the fusion process, exploring the impact of sparse noise on the fusion task. 
As a result, the proposed fusion method attains best performance.
Like other methods, our method is also limited by already registered images. 
How to register and merge is our future direction.
We will attempt to combine potential low rank representations with deep neural networks to further improve performance by utilizing the learning capabilities of the network.

\end{document}